REVIEW                                                                                                                    Open Access

# Publishing computational research - a review of infrastructures for reproducible and transparent scholarly communication

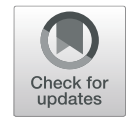

Markus Konkol[*], Daniel Nüst and Laura Goulier

## Abstract

**Background:** The trend toward open science increases the pressure on authors to provide access to the source code and data they used to compute the results reported in their scientific papers. Since sharing materials reproducibly is challenging, several projects have developed solutions to support the release of executable analyses alongside articles.

**Methods:** We reviewed 11 applications that can assist researchers in adhering to reproducibility principles. The applications were found through a literature search and interactions with the reproducible research community. An application was included in our analysis if it **(i)** was actively maintained at the time the data for this paper was collected, **(ii)** supports the publication of executable code and data, **(iii)** is connected to the scholarly publication process. By investigating the software documentation and published articles, we compared the applications across 19 criteria, such as deployment options and features that support authors in creating and readers in studying executable papers.

**Results:** From the 11 applications, eight allow publishers to self-host the system for free, whereas three provide paid services. Authors can submit an executable analysis using Jupyter Notebooks or R Markdown documents (10 applications support these formats). All approaches provide features to assist readers in studying the materials, e.g., one-click reproducible results or tools for manipulating the analysis parameters. Six applications allow for modifying materials after publication.

**Conclusions:** The applications support authors to publish reproducible research predominantly with literate programming. Concerning readers, most applications provide user interfaces to inspect and manipulate the computational analysis. The next step is to investigate the gaps identified in this review, such as the costs publishers have to expect when hosting an application, the consideration of sensitive data, and impacts on the review process.

## Background

In many scientific fields, the results of scientific articles can be based on computations, e.g., a statistical analysis implemented in R. For this type of research, publishing the used source code and data to adhere to "open reproducible research" (ORR) principles (i.e., public access to the code and data underlying the reported results [1]) seems simple. Nevertheless, several studies have concluded that papers rarely include or link to these materials [2, 3]. Reasons for that are manifold:

First, due to *technical challenges*, e.g., capturing the analyst's original computational environment, even having accessible materials does not guarantee that results can be reproduced [4, 5]. Second, many authors hesitate to share their work because publishing erroneous papers can *damage an author's reputation* [6] as well as *trust* in science [7]. These perspectives, however, overlook the fact that engaging in open practices offers some career advantages [8, 9] and can help in identifying and correcting mistakes [10, 11].

* Correspondence: m.konkol@uni-muenster.de
Institute for Geoinformatics, University of Münster, Münster, Germany

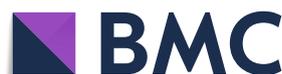

As a result of authors not including their source code and underlying data, several further problems arise. For example, reviewers *cannot verify the results*, because without the code, they are required to understand the analysis just by reading the text [12]. Hence, finding errors in results is difficult and often impossible [6], raising the question of whether the traditional research article is suitable for conveying a complex computational analysis [13]. Additionally, other researchers working in similar areas *cannot continue building on existing work* but have to collect data and implement the analysis from scratch [14]. All these issues are also to society's disadvantage, as the public cannot benefit fully from publicly funded research [15].

Funding agencies, e.g., Horizon 2020, are increasingly requiring data and software management plans as part of grant proposals. Accordingly, more journal editors are starting to make sure that author guidelines include a section on code and data availability [16, 17], and reviewers are now considering reproducibility in their decision processes [10]. Moreover, concepts and tools to package code, the computing environment, data, and the text of a research workflow (a so-called "Research Compendium" [18]) are becoming more advanced and applied. This form of publishing research allows reviewers to verify the reported results and readers to reuse the materials [19].

Nevertheless, neither the cultural and systematic developments [20] for ORR nor the existence of technologies for packaging research reproducibly can alone solve the plethora of reproducibility issues. Authors often do not know how to fulfill the requirements of funding bodies and journals, such as the Transparency and Openness Promotion (TOP) guidelines [21], or they do not have the programming skills. It is important to consider that the range of researchers' programming expertise varies from trained research software engineers to self-taught beginners. For these reasons, more and more applications have been created to support the publication of executable computational research for transparent and reproducible research. This paper aims at reviewing these applications in order to help researchers find the application that best suits their individual needs.

## Methods
### Study design
In this review study, we surveyed and compared 11 applications that assist authors in publishing reproducible research. The goal of the review was to obtain an overview of the benefits and limitations of these applications considering the challenges outlined in the previous section. We contrasted the solutions to create a set of criteria that addresses the needs of the stakeholders involved in the scholarly publication process, i.e. publishers, editors, authors, readers/reviewers, and librarians [22].

### Sample
We identified the applications during a literature search as well as through discussions at conferences[1] and workshops.[2] We included an application in our analysis if it (i) was actively maintained at the time the data for this paper was collected (5th–13th Dec 2019[3]), (ii) supports publishing executable code and data that can be inspected and reused, and (iii) is explicitly connected to the publication process. Hence, we did not consider technologies (e.g., containerization) that alone cannot support the publication process of code because further infrastructure is needed, systems that only provide access to materials (e.g., Zenodo), or workflow systems (e.g., Taverna [23]). Based on the sample criteria, we selected the following eleven applications for the review, presented in alphabetical order (see Table 1).

### Variables
In a next step, we reviewed literature to identify a set of comparison criteria (highlighted in bold in the following) relevant for the stakeholders mentioned above. According to Hrynaszkiewicz [17], publishers refrain from hosting data, raising the question of whether the applications allow (1) "free self-hosting" by the publishers. Since self-hosting might require changes to the software, we also checked whether the applications are released under an (2) "open license". Next, a proxy for assessing the stage and the reliability of an application is to check whether it is already (3) "in use". To provide an initial estimate of the application's longevity, we looked up whether the applications are (4) "grant-based", since such funds are usually temporary. Also, because using literate programming tools is a frequently mentioned recommendation for creating executable documents [35], we checked whether the applications support (5) "R Markdown" and (6) "Jupyter Notebooks". However, since researchers might have individual requirements [22], we also investigated whether the applications are (7) "extensible", meaning, for example, whether users can add a new submission format. A further relevant piece of information for authors is whether they need to (8) "upload" their materials to the application and whether (9) "copyright" is addressed explicitly in the documentation. Copyright is a main concern of authors in the context of ORR [36] and needs to be considered when it comes to reusing research materials [37]. We also checked whether (10) "sensitive data" can be shared. Based on the benefits of ORR summarized by Konkol et al. [36], we checked whether the applications provide tools to enable the (11) "discovery" of articles, (12) "inspection" of the materials, (13) "execution" of the analyses (one-click reproducible), (14) "manipulation" of parameter values, (15) "substitution" of datasets, and (16) "download" of materials. Finally, papers describing open science guidelines [21] or assessing reproducibility of published papers [3] often refer to the importance of making materials permanently available, for example, for future use and education [38]. Hence, we investigated whether it is possible to (17) "modify or delete materials after publication" and whether these materials can be (18) "shared via a DOI" or (19) "shared via a URL".

---
[1]One conference we attended: EGU General Assembly 2019 (last access of this and the following URLs: 22nd May 2020); it also had a session on open science.
[2]Workshop: eLife innovation sprint 2019, which brought together people interested in open science.
[3]A reviewer directed us to the application Authorea, which we missed in our first analysis, and the lack of pricing information. We thus collected data to address these aspects on 22nd May 2020

**Table 1** Overview of applications we included in the analysis

| Application | Description |
| --- | --- |
| Authorea | In Authorea, authors can create executable papers collaboratively. They can attach code and data to figures to make them reproducible. Authors can also directly submit to a journal and, at the same time, publish a preprint. |
| Binder | Binder creates a containerized executable environment based on a repository (e.g., on GitHub/Lab, Zenodo) including a Jupyter Notebook [24]. Readers can launch the analysis and inspect the workflow in a browser. |
| Code Ocean | Code Ocean creates "capsules" containing code, data, and the computational environment. While reading, users can execute and inspect the analysis in a separate window below the article or on Code Ocean's website [25]. |
| eLife Reproducible Document Stack (RDS) | RDS originates from the life sciences. Authors can publish executable documents based on Stencila (https://stenci.la/), an open-source editor for articles. The executable document, which contains the whole narrative and executable code snippets, is not only a supplement but the actual scientific article. |
| Galaxy | Galaxy [26] provides features tailored to use cases in the life sciences. It is a web app for developing comput. Analyses without programming expertise. Scientists can upload and analyze data using Jupyter Notebooks [27]. |
| Gigantum | Gigantum packages code, data, the computational environment, and the work history into a Git repository. Gigantum is composed of a client app for creating as well as executing analyses locally and a cloud-based infrastructure for sharing computations and collaborating with peers. |
| Manuscripts | Manuscripts is an online tool for writing executable documents collaboratively based on the concept of literate programming, but featuring a "What you see is what you get" user interface. The runtime environment of the author is, however, not considered. |
| o2r | o2r [22] originates from the geosciences and addresses publishers who want to extend their infrastructure via a reproducibility service during the process of paper submission [28]. Authors can create interactive figures, allowing readers to change model parameters using a slider [29]. |
| REANA | REANA [4, 30] originates from particle physics and provides a specification for capturing data, code, and the comput. Environment. Based on this structure and manually created configuration files, REANA provides command line interface (CLI) commands to run large analyses on a remote REANA cloud. |
| ReproZip | ReproZip [31, 32] provides a set of CLI commands for encapsulating data, code, and the computational environment. Users can execute the resulting bundle on a server provided by ReproZip [33] or locally on different systems. |
| Whole Tale | With Whole Tale [34], authors can create so-called "Tales" that combine narrative, data, code, and the computational environment. Readers can inspect the materials and execute the analysis in the original environment. |

### Data collection
Based on the comparison criteria, two authors collected information iteratively by investigating the project websites, applications, GitHub/Lab repositories, scientific articles, and blog posts. In the first iteration, one author of this paper gathered information on the applications one by one and took screenshots. In the second iteration, another author independently checked the information collected in the first iteration. Conflicts concerning the data were resolved through discussion amongst all authors. In order to give the scientific community, particularly the developers of the considered applications, the opportunity to comment on the analysis, we published a preprint [39] of the paper three months before submission. All collected data is available in the supplement (see Availability of data and materials). Thus, it is possible to continue the work in this paper as a "rolling review". Since some sources (e.g., documentation) were not scientific articles, we also attached links to and screenshots of the original information.

### Results
Table 2 summarizes aspects relevant for publishers, editors, authors, readers, and librarians.

**Table 2** Overview of which application supports the corresponding criteria. (N/D = no data)

|  | Authorea | Binder | Code Ocean | eLife RDS | Galaxy | Gigantum | Manuscripts | o2r | REANA | Repro Zip | Whole Tale |
|---|---|---|---|---|---|---|---|---|---|---|---|
| Free self-hosting | − | + | − | +* | + | − | + | + | +* | + | + |
| Open license | − | + | − | + | + | +/− | + | + | + | + | + |
| In use | in use [40] | in use [2] | in use [41] | in use [42] | in use [43] | − | − | − | in use [44] | in use [31] | − |
| Grant-based | − | + | − | + | + | − | N/D | + | + | + | + |
| R Markdown | − | + | + | + | − | + | − | + | − | − | + |
| Jupyter Notebooks | + | + | + | + | + | + | − | − | + | + | + |
| Extensible | − | + | + | + | + | − | − | − | + | + | + |
| Upload | + | + | + | − | + | − | + | + | − | − | + |
| Copyright | + | N/D | + | N/D | + | + | N/D | + | N/D | N/D | + |
| Sensitive data | − | − | − | − | − | − | − | − | − | − | − |
| Discovery | + | − | + | + | + | − | − | + | − | − | + |
| Inspection | + | + | + | + | + | + | + | + | − | − | + |
| Execution | + | + | + | + | + | + | + | + | + | + | + |
| Manipulation | + | + | + | + | + | + | + | + | + | + | + |
| Substitution | − | − | − | − | − | − | − | + | − | + | − |
| Download | + | + | + | + | + | + | + | + | − | + | + |
| Modify/Delete after publishing | − | + | − | − | + | + | + | − | + | + | − |
| Shared via DOI | + | − | + | + | − | − | − | − | − | − | + |
| Shared via URL | + | + | + | + | + | + | + | + | − | + | − |

From the eleven applications, eight allow self-hosting for free. eLife RDS and REANA (in Table 2 marked by *) require deployment, since no free online version exists. Eight applications are released under an open license, and Gigantum has only published the client tool under an open license. The open source applications that have an online version running can be used by researchers for free. The three commercial providers, namely Authorea, Code Ocean, and Gigantum, provide the service in exchange for payment but they also offer free subscriptions with limited features and resources (i.e. storage and computation time).

In total, seven applications are already in use, as shown by the example papers with reproducible workflows. Seven applications currently receive funding from public or private organizations.[4]

Ten applications support literate programming, e.g., R Markdown or Jupyter Notebooks; the Manuscripts application supports Markdown but also code execution via embedded Jupyter Notebooks. Seven applications are extensible and can be configured to support further programming languages. Except for Code Ocean, which also supports MATLAB and Stata, all applications only support non-proprietary programming languages.

Seven applications require authors to create their projects online, whereas eLife's RDS (based on Stencila), REANA, and ReproZip allow local usage. Researchers can also work locally with Gigantum, but they then need to synchronize with the online service to access all features.

Regarding copyright, we could not find explicit information on assigning copyright for research materials in five applications. Whole Tale and Gigantum only allow open licenses, whereas Code Ocean, Galaxy, and o2r encourage them. We could not find information on sensitive data in any of the applications.

From the eleven applications, six provide a keyword-based search for papers, of which o2r provides a spatio-temporal search combined with thematic properties, such as libraries used in the code. Five applications embed a programming environment (e.g., JupyterLab, RStudio) for inspecting code and data, whereas four provide their own user interface (UI).

All applications provide support for executing the analysis. REANA projects are executed via the CLI in a remote REANA cloud, and this also applies to ReproZip, which in addition to a remote cloud also provides a ReproServer for executing code online. Gigantum's local client allows users to run code in the browser. The remaining applications allow users to execute the analysis in a browser on a remote server.

---

[4]Further details on funding are available in the supplement.

Each application allows users to manipulate the code and rerun it based on a new parameter. Most commonly, users can directly manipulate the code in the browser (8 applications provide this option). In REANA and ReproZip, users can pass new parameter values via the CLI, whereas the o2r platform enables authors to configure UI widgets that allow reviewers/readers to manipulate parameter values interactively, e.g., by using a slider to change a model parameter within a defined range. Features for substituting the input datasets used in an analysis are provided by o2r and ReproZip.

While ten applications provide a feature for downloading materials, REANA projects need to be stored on a third-party service to be downloadable.

Overall, six applications allow users to modify/delete materials after publication. In Binder, REANA, and ReproZip, modifying/deleting content is possible if the research materials are stored on GitHub/Lab, but not when they are stored on Zenodo. Authorea, Code Ocean, eLife RDS, and Whole Tale assign DOIs to published content, ensuring long-term availability and making it impossible to edit these after publication. In eLife's RDS, the article is composed of text and code; thus, deleting it is equivalent to withdrawing a paper.

## Discussion

Several developers have created applications for publishing computational research. One might think the applications, since they all strive for the same overall goal, resemble each other. However, we showed in this paper that the applications address different issues and needs, which increases the chances that stakeholders will find an application that best suits their individual requirements. The review can be used by the various stakeholders in different ways: Publishers who want to comply with reproducibility principles may use it to decide for a certain application, editors and program committees may use it when planning to include code review in their review process [45], applicants designing data and software management plans may use it when writing their funding proposals, and authors who are searching for tools to disseminate their work in a convincing, sustainable, and transparent manner may also find it valuable. In addition to these stakeholders, we also considered librarians, who are tasked with aspects related to the preservation and long-term accessibility of research materials. Given the variety of the stakeholders and their considerations, it is difficult to determine the best application or objectively provide a ranking. Identifying the ideal application strongly depends on the needs and goals of the stakeholders.

### Hosting of the applications

Publishers need to decide whether they want to host an infrastructure by themselves or engage a provider. Applications exist for both approaches, though the majority of them can be self-hosted. Some of the self-hosting solutions are also available as online versions, but it should be considered that these have limited resources regarding storage and processing power. Moreover, it is difficult to estimate when the projects will expire. If the applications are grant-based, they might receive follow-up funding, but this depends on the projects' success and whether they aim at carrying out research (which might come with many changes to the software) or developing a scalable and sustainable platform. As public information on all funding levels and grant durations are too uncertain and incomplete to be included in the analysis, we refrained from drawing concrete conclusions in terms of longevity and how likely they will exist in the next years.

All self-hosting solutions have an open license, allowing operators to host their own service as well as modify the software according to their individual needs and styles. A further advantage of self-hosting is the mitigation of risks regarding vendor lock-in. However, hosting one's own service means that publishers also have to provide the required technological resources and personnel. It remains unclear what kinds of costs publishers will have to expect when hosting a platform and incorporating it into their publishing infrastructure. The final costs strongly depend on the number of views, execution attempts, workflow sizes, and ease of integration into technical systems. These parameters differ between use cases and could be used as measures for future research, e.g., on stress tests. Therefore, the metrics of existing publications might provide the first ways to calculate the required resources. While the Binder instance MyBinder.org published an initial estimate regarding costs,[5] further data from other services would help to calculate costs more specifically. Moreover, it would be interesting to see usage statistics showing how often the services are used, for example, by authors, readers, and reviewers, albeit this transparency is only realistic for non-profit projects. Nevertheless, since reproducibility studies are rarely successful [5], using these services seems to be uncommon.

A further criterion we investigated was whether the applications are in use. While applications in use offer initial evidence that they work, it might take more effort to adjust them to fit a publisher's infrastructure. In contrast, beta applications can adjust their features without worrying about running instances and backwards compatibility, but the deployment of such applications might reveal new issues.

---

[5]MyBinder costs: https://mybinder.org/v2/gh/jupyterhub/binder-billing/master?urlpath=lab/tree/analyze_data.ipynb

### Creating executable analyses

Regarding submission formats, there is a trend toward literate programming approaches. Most applications either support Jupyter Notebooks or R Markdown, which both have proven to support reproducibility [46]. However, some journals and publishers rely on different formats, e.g., LaTeX. Transformations to other document types are often cumbersome and adapting author requirements can be a lengthy process. Hence, it might be easier to have reproducible documents as a supplement, potentially for a transition period, until researchers have adjusted their workflows and executable documents are widely accepted. Nevertheless, eLife's RDS has already shown that combining executable code with narrative in a scientific article is possible today and comes with advantages related to communicating scientific results. For example, readers can, while studying the text, also manipulate the analysis. A limitation in this context is related to the peer-review process. All applications require an account for creating reproducible results, and since the name of the creator is usually visible, a double-blind review is not guaranteed. However, access to code and data is particularly important for reviewers who need to recommend acceptance or rejection of a submission. One solution might be to create anonymous versions of the materials, as is possible with Open Science Framework,[6] or to adopt an open peer-review process.

A further critical issue is that not all applications address copyright in their documentation. Those that do either require or encourage open licenses, which is a recommendation mentioned frequently in papers discussing reproducibility guidelines [21, 37]. Hence, the platforms should inform users about licenses, e.g., by referring to existing advising resources (e.g., https://choosealicense.com/). Licensing is important to enable reusability and, thus, is ideally assigned to code, data, and text separately, as is done, for example, by Authorea. Computational reproducibility is challenging also because of sensitive data. None of the applications address this issue, but platforms allowing self-hosting can be combined with existing solutions, such as involving trustworthy authorities [47] and cloud-based data enclaves [48].

---

[6]Anonymized links: https://help.osf.io/hc/en-us/articles/360019930333-Create-a-View-only-Link-for-a-Project

### Studying reproducible research

Being able to reproduce the computational results in a paper is a clear benefit, but open reproducible research comes with a number of further incentives [20]. Concerning the discovery of papers, most search tools provided by the applications do not take full advantage of the information contained in code and data files, e.g., spatiotemporal properties. Instead, they either only provide a keyword-based search or no search at all. For inspecting materials, most solutions either provide their own UI or integrate a development environment, e.g., JupyterLab. In both cases, users can directly access, manipulate, and reuse the code. However, readers (including experienced programmers) might still find it challenging to understand complex code scripts. Moreover, identifying specific parameters buried in the code and finding out how to change these can be a daunting task. The concept of nano-publications [49] or bindings [29] might help to solve these issues. A further need in this context is a UI for comparing original and manipulated figures, since differences in the figure after changing parameters might be difficult to spot. Most applications do not provide any support for substituting research components, e.g., by other input datasets, which might be due to the plethora of complex interoperability issues with respect to data formats or column names in tabular data. Only ReproZip [32] and o2r [36] provide basic means to substitute input datasets, yet they require users to ensure compatibility.

Researchers who are writing or studying computational research articles might be programming beginners or experts. Hence, while the learning curve may be either shallow or steep, it is present in any case. Although the applications are well documented, programming novices in particular might need to invest effort at the beginning of use. For example, they would need to learn how to write R Markdown documents and create configuration files manually. Some of the creation steps might be automated, but this usually comes at the cost of flexibility. The learning curve not only exists for authors but also for consumers, particularly reviewers who need to verify the results and those who want to build upon the materials. Nevertheless, such an effort only needs to be invested once and will eventually result in more convincing and transparent research.

### Sharing computational research

The state of the research materials is an issue when it comes to publication. While some applications fix the state of the research materials by assigning a DOI and archiving a snapshot, others allow changing and deleting them. This is a disadvantage with respect to reproducibility since verifiability and accessibility are lost. In addition, if self-hosting is not possible, the computational analysis of an article will be executable only as long as the project and its infrastructure exist; this dependence is a crucial aspect with respect to archiving. However, this issue can be mitigated if researchers "go the extra mile" and also publish their materials in long-term repositories in addition to an executable version using one of the applications.

A further dependence is the technology underlying the infrastructure. For example, without the Docker container runtime, the captured computing environment might not work even though it remains human readable [50]. This is also true for source code scripts, which are plain text files and, thus, can be opened using any editor, even if they cannot be compiled and executed. These examples demonstrate the importance of using open and text-based file formats instead of proprietary and binary file formats in science.

### Limitations

This work is subject to a number of limitations. The scope of this review is narrow and does not cover all applications that are connected with computational research (e.g., workflow systems, such as Taverna [23]). Also, we have no access to publishers' actual systems, preventing us from being able to evaluate the usability of APIs and documentation and how easy they can be incorporated into existing infrastructures. In addition, this review is a snapshot of the highly dynamic area of publishing infrastructures. Hence, the information might become outdated quickly, e.g., an application might extend the set of functionalities or be discontinued. Still, reviewing the current state of the landscape to reflect on available options might be helpful for researchers. Furthermore, the properties we investigated in this survey do not cover all possible aspects and discipline-specific needs, but, nevertheless, stakeholders requiring more information can use the overview as a starting point for further research. Also, we collected and interpreted the data ourselves and did not contact the application developers, which might have increased the accuracy of the data. Finally, our evaluation only considered documented features. However, programmers with sufficient expertise can build upon the open source applications and implement missing features.

### Conclusions

In this review, we compared eleven applications in order to identify their benefits and limitations for assisting researchers to publish and study open reproducible research. Our findings show that publishers have the choice between using provided services or self-hosting solutions, but more data is needed to estimate the costs for publishers to maintain their own infrastructure. The review revealed a trend towards literate programming approaches as well as tools for reviewers and readers, e.g., for inspecting an analysis or manipulating the assumptions underlying the analysis. We found that being able to change the materials after publication might result in conflicts between the version referred to in an article and the available version, which might have been changed since the article was first published. In addition to investigating these issues, the next step is to examine how using an application affects a reviewer's decision and how much additional effort is needed to study the materials.

#### Abbreviations
CLI : Command line interface; DOI : Digital Object Identifier; o2r : Opening reproducible research (project); ORR : Open reproducible researchers; REANA : Reproducible Analysis (project); RDS : Reproducibility Document Stack; UI : User Interface

#### Acknowledgements
We thank Timothy Errington and Mario Malicki for their reviews, Vicky Steeves for a helpful discussion on the article preprint [39], and Celeste Brennecka for proofreading.

#### Authors' contributions
Markus Konkol wrote the paper, collected the data, and conceptualized the analysis. Daniel Nüst wrote the paper. Laura Goullier collected data and wrote the paper. All authors discussed the results and approved the final manuscript.

#### Funding
This work is supported by the project Opening Reproducible Research 2 (https://www.uni-muenster.de/forschungaz/project/12343) funded by the German Research Foundation (DFG) under project numbers KR 3930/8–1; TR 864/12–1; PE 1632/17–1. The funders had no role in study design, data collection and analysis, decision to publish, or preparation of the manuscript.

#### Availability of data and materials
The data is openly available on Zenodo: https://doi.org/10.5281/zenodo.3562269. The repository includes a list of all applications we looked at and, for excluded applications, the reasons for exclusion.

#### Ethics approval and consent to participate
Not applicable.

#### Consent for publication
Not applicable.

#### Competing interests
The authors of this paper are members of the o2r project that was also discussed in this paper (http://o2r.info/).

Received: 5 March 2020 Accepted: 24 June 2020
Published online: 14 July 2020

## Publisher's Note